\documentclass[12pt]{article}

\usepackage{times}

\usepackage[sort&compress,round,comma,numbers]{natbib}
\bibpunct{(}{)}{,}{n}{;}{,}

\usepackage{graphicx}

\newcommand{\be}[1]{\showlabel{#1}\begin{equation}\label{#1}}
\newcommand{\ee}{\end{equation}}

\newcommand{\showlabel}[1]{}  
\newcommand{\randnotiz}[1]{}

\newcommand{\etal}{\emph{et al.}}	        

\begin{document}

\pagestyle{myheadings}
\markright{PufX and photosynthetic efficiency}

\title{Form follows function --- how PufX increases the efficiency of the
light-harvesting complexes of \emph{Rhodobacter sphaeroides}}

\author{Tiham\'er Geyer$^\dag$\\
		Zentrum f\"ur Bioinformatik, 
		Universit\"at des Saarlandes, \\
		D--66041 Saarbr\"ucken, Germany}

\date{Feb. 9, 2007}

\maketitle

\begin{abstract}
	Some species of purple bacteria as, e.g., \emph{Rhodobacter
	sphaeroides} contain the protein PufX. Concurrently, the light
	harvesting complexes 1 (LH1) form dimers of open rings. In mutants
	without PufX,  the LH1s are closed rings and photosynthesis
	breaks down, because the ubiquinone exchange at the reaction
	center is blocked. Thus, PufX is regarded essential for
	quinone exchange.
	
	In contrast to this view, which implicitly treats the LH1s as
	obstacles to photosynthesis, we propose that the primary purpose
	of PufX is to improve the efficiency of light harvesting by
	inducing the LH1 dimerization. Calculations with a dipole model,
	which compare the photosynthetic efficiency of various
	configurations of monomeric and dimeric core complexes, show that
	the dimer can absorb photons directly into the RC about 30\% more
	efficient, when related to the number of bacteriochlorophylls, but
	that the performance of the more sophisticated dimeric LH1 antenna
	degrades faster with structural perturbations. The calculations
	predict an optimal orientation of the reaction centers relative to
	the LH1 dimer, which agrees well with the experimentally found
	configuration.
	
	For the increased required rigidity of the dimer additional
	modifications of the LH1 subunits are necessary, which would lead
	to the observed ubiquinone blockage, when PufX is missing.
\end{abstract}

{\setlength{\parindent}{0cm}


%

\bigskip\bigskip

$^\dag$Corresponding author. Address: Zentrum f\"{u}r Bioinformatik,
Universit\"at des Saarlandes, Geb. C7.1,
Postfach 151150,
D--66041 Saarbr\"ucken, Germany \\
Email: tihamer.geyer@bioinformatik.uni-saarland.de
}

\clearpage

\section{Introduction}

Purple bacteria as, e.g., \emph{Rhodobacter (Rb.) sphaeroides} can
live on photosynthesis. In this conversion of light into chemical
energy, four transmembrane proteins and two electron carriers are
involved. It is initiated by photons, which are absorbed in the
bacteriochlorophylls (Bchls) of the light harvesting complexes. Their
energy is passed on to the special pair Bchls of the RCs. From there,
an excited elctron is translocated through the RC onto a bound
ubiquinone. Loaded with a second electron and two protons, the reduced
quinone unbinds from the RC and delivers its freight to the cytochrome
$bc_1$ complex. From there, the electrons are returned to the RC by
two cytochrome $c_2$ and the protons are released to the periplasm.
The resulting proton gradient across the membrane is used by the
F$_0$-F$_1$-ATP synthase to synthesize ATP. For more details see e.g.,
\cite{HU02, GEY06b}.

As can be seen on recent atomic force microscopy (AFM) and cryo
electron microscopy (EM) images, the photosynthetic membranes of
purple bacteria are crowded with the ring shaped LH1s with their
embedded RCs and with the auxilliary LH2
\cite{BAH04a,SIE04,SCH04b,SCH05}. In most purple bacteria, the primary
LH1 form closed rings of 16 dimeric subunits \cite{JAM02}. Each of the
subunits consists of two transmembrane helices and two
bacteriochlorophylls (Bchl), which are the functionally active parts
of the LHCs. In the center of each LH1 ring sits an RC. This assembly
of an LH1 with its embedded RC is called a core complex \cite{FOT04}.
The smaller LH2 are rings of eight or nine subunits only, depending on
the species \cite{KOE96, SCH04b}.

In some species as, e.g., \emph{Rb. sphaeroides} or \emph{Rb.
capsulatus}, an additional small protein PufX is present and the
core complexes are dimers of two RCs and two incomplete LH1s of 12 to
13 subunits, each \cite{JUN99, FRA99, SCH04a, SIE04, BAH04a}.
PufX lacking mutants of \emph{Rb. sphaeroides} have closed monomeric
LH1 rings and are not able to live on photosynthesis. This deficiency,
as shown experimentally, stems from the closed LH1 rings, which slow
down the quinone exchange at the RCs to a crawl, such that the RCs are
effectively shut off \cite{BAR95a, BAR95b}. Mutants, where the LH1s
are missing, do not require PufX for photosynthetic growth
\cite{MCG94}. Thus, the current hypothesis about the purpose of PufX
is that it opens up the LH1 ring to allow the quinones to access the
RC. This hypothesis is confirmed by the latest EM images of the LH1/RC
dimers, where the RCs are oriented such that the gap in the LH1
structure is in front of the quinone binding pockets of the RCs
\cite{QIA05}.

However, this hypothesis about the purpose of PufX does not explain,
why most other species of purple bacteria happily live on
photosynthesis without PufX and with closed LH1 rings. But there is
also a conceptual difficulty with this hypothesis: by making PufX
responsible for opening the LH1 for quinone access, the LH1 are
implicitly seen as obstacles to efficient photosynthesis, not as an
integral and important part of it. The function of the LHCs is to
capture photons, they are the antennae of the RCs. The central
objective for them is, consequently, to achieve a maximal absorption
cross section for photons with a given limited number of Bchls and
also to feed the captured photons into the RCs with the least possible
loss. We therefore put forth a different hypothesis and propose that
the primary purpose of PufX --- together with some associated
modifications --- is to increase the absorption efficiency of the core
complexes by inducing their dimerization. Then, allowing for a direct
access of the quinones to the RCs becomes necessary because of these
modifications to the then open LH1s.

To support our hypothesis, we first present calculations of the
absorption properties of LH1/RC core complexes, which compare the
monomeric type without PufX to the dimeric PufX$^+$ configuration.
These calculations, which are based on a simple dipole model of the
Bchl arrays \cite{HU97}, show that the dimeric configuration can absorb photons directly
into the RC at least as good as the monomer, though it has less Bchls.
We also find that for this the dimer has to be structurally more
rigid. The same advantage is found for monomeric core complexes with
an open LH1 ring, a setup, which is found in \emph{Rhodopseudomonas}
(\emph{Rps.}) \emph{palustris} \cite{ROS03}. From these findings and
from recent experimental results, we then argue that the observed
blocking of the quinone access to the RCs in PufX$^-$ mutants is a
consequence of the increased rigidity of the core complexes from these
bacteria, a rigidity, which is necessary to stabilize the dimeric
LH1s.

In this publication, which focuses on the differences between the
monomeric and the dimeric LH1s, we do not consider the auxilliary LH2s
because (i) they are not affected by the presence or absence of PufX
and (ii) because their coupling to the core complexes is indirect and
relatively slow. In the big picture, their contribution to
photosynthesis is to the also neglected non-resonant energy transfer
from the LH1s to the RCs.

\section{Methods}

\subsection{Dipole model of the core complex}

The calculations of the absorption properties of the different core
complex configurations are based on a dipole model introduced by Hu
\etal{} \cite{HU98b, HU97} (also see \cite{COR98, SRO06}). There, the
positions and orientations of the RC Bchl dipoles had been determined
from the crystal structure \cite{KOE96}, while the dipoles of the
monomeric LH1 ring were derived from a reconstruction.

Our objective was to compare different core complex configurations
from the same species. These are different on a large scale, but can
be expected to have the same local environment of the Bchls and the
same next neighbor distances. Thus, the same parameters were used for
all configurations.

The Bchl positions in the dimeric LH1 were determined by visually
fitting two three-quarter rings of the LH1 monomer symmetrically into
an EM map of the LH1 dimer. The absorption properties are only
minimally sensitive to the exact positions and orientations of the two
dimer halves. The two RCs of the dimeric core complex were placed
symmetrically into the respective centers of the two halves of the LH1
dimer. For the calculations, the rotation angle of the RCs with
respect to their initial orientation, $\Phi_{RC}$, was treated as a
free parameter.

The open monomeric core complexes were constructed from the closed
monomer by removing adjacent LH1 Bchl dipoles. Here again, the
orientation of the RC with respect to the LH1 remainder was treated as
a free parameter. The shape of the initially circular LH1 ring was not
distorted, also the RC was always put at the center of the original
circle, even though the open rings found in \emph{Rps. palustris} have
an unsymmetric elliptical shape.

\begin{figure}[t]
	\begin{center}
		\includegraphics[]{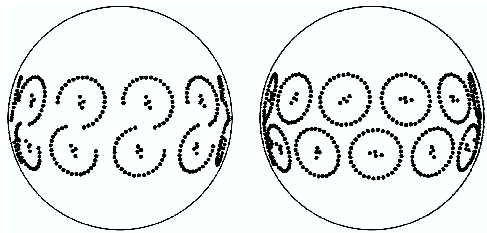}
	\end{center}
	\caption{Sketch of the arrays of core complexes on a vesicle of 50
	nm diameter that were used for our calculations. The dots denote
	the positions of the Bchls of the 11 Z-shaped dimers (left panel)
	and of the 24 closed monomers (right panel). In the dimeric setup
	the RCs are shown in their most efficient orientation, while in
	the monomeric configuration they are oriented randomly.}
	\label{fig:OnTheVesicle}
\end{figure}

Arrays were built from the closed monomeric and the dimeric core
complexes and placed onto a spherical vesicle of 50 nm diameter.
According to AFM images \cite{BAH04a} and our reconstruction of a
chromatophore vesicle \cite{GEY06a}, the dimers were assembled as a
chain with each unit rotated by 10 degrees clockwise. To fit onto the
vesicle, the dimers had to be bent at their joint by 26$^\circ$. The
vesicle was large enough for 11 dimeric core complexes with a total of
616 Bchls. For comparison, a similar setup on a vesicle was
constructed with monomeric core complexes, too, where 24 monomers with
their 864 Bchls were arranged alternatingly in two rows parallel to
the equator of the vesicle. Both configurations are shown in figure
\ref{fig:OnTheVesicle}.

\subsection{Total absorption cross section and photosynthetic efficiency}

The eigenstates $|\Phi_n \rangle = \sum\, a_{ni}|i\rangle$ of an array
of dipoles and their absorption cross sections, determined by their
oscillator strengths $|f_n^2|$, were calculated as explained in
\cite{HU97} with a Hamiltonian from all Bchls of the given
configuration.

The total absorption cross section of a certain configuration is $\sum
f_n^2 = N$ according to the dipole summation rule. It states that no
absorption cross section is lost by coupling the $N$ dipoles.
Consequently, the total absorption cross section is not a meaningful
measure for the efficiency of a given core complex configuration,
because it is solely determined by the number of Bchls.

Absorbing light in an LHC is only the very first step of
photosynthesis. The absorbed photons then have to be transferred to
the special pair Bchls of the RCs in order to trigger a charge
separation. Between the absorption and the charge separation, the
energy of the photon can be lost due to thermal relaxation. This
efficiency degrading loss process becomes the more important, the
longer the electronic excitation takes to travel from the Bchls of the
LHCs to the special pair of the RC. Consequently, the most lossless
transfer would be an absorption of the photon directly into the
special pair.

To describe how direct a given state absorbs photons into the special
pair, we introduce the photosynthetic cross section $\sigma_n$ of an
eigenstates $|\Phi_n\rangle$ of a given core complex configuration. It
is the product of its absorption cross section $f^2_n$ and of the
probability $S_n$ that one of the special pair Bchls is excited in
this state. $S_n$ is calculated from the incoherent sum of the weights
$|a_{ni}|^2$ of the special pair (SP) Bchls:
\be{eq:photoCS}
	\sigma_n = f^2_n\;S_n = f^2_n \; \sum_{SP} |a_{ni}|^2
\ee
From this state specific absorption into the RC we define the total
photosynthetic cross section $\Sigma$ of a given configuration as
$\Sigma = \sum \sigma_n\,$. Obviously, for $\Sigma$ there is no
summation rule. Different configurations with the same number of Bchls
may have different total photosynthetic cross sections. From the two
cross sections $\sum f_n^2 = N$ and $\Sigma$, the photosynthetic
efficiency $\eta$ is introduced as $\eta = \frac{\Sigma}{N}$. This
efficiency can either be interpreted as the fraction of absorbed
photons that is directly available to induce a charge separation in
the RC, or as the fraction of the Bchls that couples directly to the
RC Bchls.

\subsection{Thermal disorder}

Thermal disorder of the Bchls modifies their positions and
orientations and, by this, their site energies and coupling
parameters, and finally the photosynthetic cross section. As only
direct, instantaneous photon capture into the special pair Bchls is
considered, we can assume that the thermal fluctuations are much
slower than the actual photon absorption. The resulting quasistatic
deformations of the core complex, which consequently also include
static spatial deformations, are captured in the effective Hamiltonian
model by a random perturbation of the site energies and of the
coupling terms.

To investigate the stability of $\Sigma$ against thermal fluctuations
and spatial deformations, the off--diagonal entries of the Hamiltonian
that characterize the interactions and the diagonal entries for the
site energies were independently multiplied by random numbers drawn
from a Gaussian distribution centered around 1 with a relative width
of $\Delta E/E$ of up to 12\%. The distribution was modified
such that the product of all random numbers is 1, i.e., that the
fluctuations do not introduce an energy shift. Perturbing only
the interactions or the site energies leads to the same behavior of
$\Sigma$, however, for the same effect the interaction terms had to be
perturbed about four times as strong as the site energies. To achieve
stable average values for $\Sigma$, the calculations were repeated 200
times for every chosen $\Delta E/E$.

\section{Results}

\subsection{Closed monomeric core complexes}

The benchmark configuration of the closed monomeric core complex,
which is found in most purple bacteria, consists of the 16--unit LH1
ring with two Bchls each, and an embedded RC. In the dipole model, the
empty LH1 ring with its essentially circular symmetry has two
degenerate states with orthogonal dipole moments, absorbing at a
wavelength of 875 nm. These two states, with $f^2_{2,3} = 15.7$ each,
carry most of the total oscillator strength of $\sum f^2_n = 32$
\cite{HU97} (All cross sections are given in units of the dipole
moment of the $S_y$ transition of a Bchl).

With th RC inside the LH1 ring, the circular symmetry is broken. From
one of the two main LH1 states and the RC groundstate two hybrid
LH1-RC states emerge with oscillator strengths of $f^2_2 = 13.0$ and
$f^2_4=5.82$ and energies corresponding to wavelengths of 876 nm and
864 nm, respectively. The other LH1 state remains unchanged, as its
dipole moment is perpendicular to the RC dipoles. These three states
2, 3, and 4 together are responsible for 96\% of the total absorption
cross section. Their respective photosynthetic cross sections, i.e.,
their cross section for absorption directly into the special pair
Bchls of the RC, are $\sigma_2=0.82$, $\sigma_3=8\times 10^{-5}$, and
$\sigma_4=4.25$: photosynthesis runs on the LH1/RC hybrid states 2 and
4, while the probility to induce a charge transfer in the RC is
negligible in state 3 with its dipole moment orthogonal to the RC
dipoles. Summing up all $\sigma_n$ results in a total photosynthetic
cross section of $\Sigma_M = 5.22$ for the closed monomeric core
complex and a corresponding efficiency of $\eta_{M} = 0.145$:
effectively one out of every seven of the 36 Bchls contributes
\emph{directly} to photosynthesis. The other 85\% of the absorbed
light have to be handled by higher order transitions between the
states, by energy downconversion processes, or they are directly
dissipated as heat.

\subsection{Open dimeric core complexes}

While in the dipole model the closed LH1 ring has a circular symmetry,
the Z--shaped LH1 dimer only has a twofold symmetry axis and its
eigenstates are not degenerate. Its absorption spectrum is dominated
by the three low lying states 1, 3, and 5, which absorb at 882, 875,
and 866 nm, respectively. Their oscillator strengths are $f^2_1=10.0$,
$f^2_3=27.6$, and $f^2_5=6.97$, i.e., together they account for 94\%
of the total absorption cross section. Interestingly, state 5 has an
energy very close to the RC groundstate at 865 nm. It can be expected
that this state will couple very well to the RCs.

\begin{figure}[t]
	\begin{center}
		\includegraphics[]{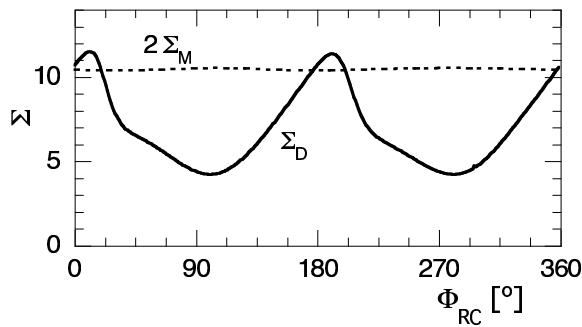}
	\end{center}
	\caption{Total photosynthetic cross section $\Sigma_D(\Phi_{RC})$
	of the LH1 dimer with its two RCs (solid curve), compared to two
	times $\Sigma_{M}$ of a single RC in a closed monomeric LH1
	(broken curve). The angle $\Phi_{RC}$ denotes the orientation of
	the RCs relative to their initial positions. Note that two
	monomeric core complexes together contain 72 Bchls, while the
	dimeric core complex only has 56 Bchls.}
	\label{fig:DimerRotateRCs}
\end{figure}

With the RCs inserted into the LH1 dimer, the energies and oscillator
strengths, i.e., the absorption spectrum of the LH1/RC combination
states, vary only little with the rotation angle $\Phi_{RC}$ of the
RCs. However, the photosynthetic efficiency, which is determined by
the coupling between the LH1 and the RCs, is very sensitive to
$\Phi_{RC}$.

Figure \ref{fig:DimerRotateRCs} compares $\Sigma_D(\Phi_{RC})$ of the
dimeric core complex to twice $\Sigma_M$ of the monomer. As expected
from the circular symmetry of the monomer, $\Sigma_M$ is constant for
all orientations of the RC, while for the dimer there are two
pronounced maxima spaced 180$^\circ$ apart. At these maxima, the
dimeric core complex with its 56 Bchls can absorb photons directly
into the RCs with a photosynthetic cross section of $\Sigma_D = 11.5$.
Two independent monomers with their 72 Bchls, present a cross section
of only $2\Sigma_M = 10.4$. Consequently, for optimal orientation of
the RCs, the efficiency of the dimer of $\eta_D = 0.21$ is about 30\%
higher than that of the monomer of $\eta_M = 0.15$.

\begin{figure}[t]
	\begin{center}
		\includegraphics[]{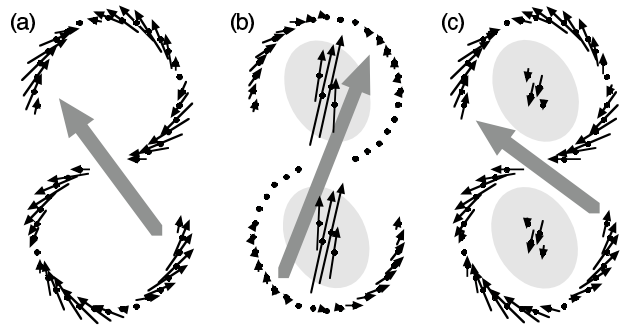}
	\end{center}
	\caption{Sketch of state 5 of the empty LH1 dimer (panel (a)) and
	of states 5 and 7 of the dimeric core complex with optimally
	oriented RCs (panels (b) and (c), respectively). These two states
	are formed from state 5 of the empty dimer and the RC groundstate.
	State 5 (panel (b)) is the photosynthetically active state, while
	in state 7 the coupling between LH1 and RCs is negligible. The
	dots denote the positions of the dipoles in the membrane plane as
	seen from the cytoplasmic side, the black arrows give the
	direction of each Bchl dipole and, via their length, their weights
	$|a_{ni}|^2$ in the respective state. The total dipole moment of
	the states is denoted by the gray arrows. The position of the RCs,
	as indicated by the shaded regions, compares well to the
	reconstruction by Qian \etal{} \cite{QIA05}.}
	\label{fig:DimerDipoles}
\end{figure}

The orientation of the RCs and the individual weights $|a_{ni}|^2$ of
the dipoles at the optimal orientation of the RCs are sketched in
figure \ref{fig:DimerDipoles} for the two most important states. For 
comparison, panel
(a) shows state 5 of the empty dimer, which absorbs at 866 nm. This
state couples to the RC groundstate to form the two states 5 and 7
of the dimeric core complex shown in panels (b) and (c). Both states
absorb at about 865 nm, but with photosynthetic cross sections of
$\sigma_5 = 9.66$ and $\sigma_7 = 0.2$, respectively, i.e, they form
one ``photosynthetic'' and one ``anti--photosynthetic'' state.

Interestingly, the orientation of the RCs, as indicated in figure
\ref{fig:DimerDipoles}, corresponds well to the orientation found in
the reconstruction by Qian \etal{} \cite{QIA05}. Thus, not only the
access for the quinones to and from the RCs is possible through the
gap in the LH1, but it is also the configuration, in which the
coupling between the LH1 and the special pair Bchls of the RCs is most
efficient.

One can see from figure \ref{fig:DimerRotateRCs} that for
bacteria with dimeric core complexes it makes a huge difference,
whether all RCs are oriented optimally with respect to their LH1
antenna or whether they are oriented randomly. Thus, in PufX
containing species there should be some mechanism to lock the
orientation of the RCs inside the LH1, a feature which is not required
with the symmetric closed LH1 monomers.

\subsection{Thermal disorder and structural stability}

Next we have to investigate the performance of the core complexes 
under more realistic conditions than in the fixed setup of zero 
temperature used above.

\begin{figure}[t]
	\begin{center}
		\includegraphics[]{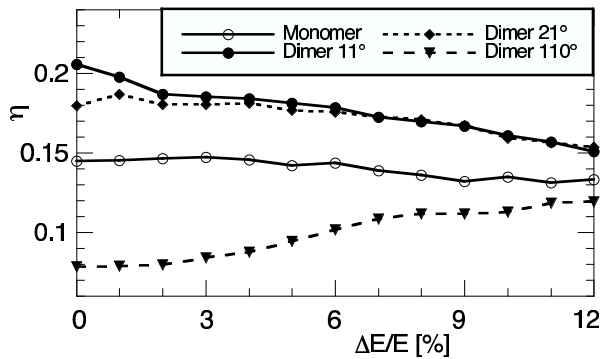}
	\end{center}
	\caption{Photosynthetic efficiency $\eta$ vs. thermal
	perturbation, indicated by the relative width of the distribution
	of the interaction strengths, $\Delta E/E$, for the monomer and
	the dimer with three different orientations of the RCs (cf. figure
	\ref{fig:DimerRotateRCs}). The lines connecting the data points
	serve as a guide to the eye. For further explanations see text.}
	\label{fig:Rauschen}
\end{figure}

Figure \ref{fig:Rauschen} shows, how the photosynthetic efficiency
$\eta$ decreases with increasing thermal disorder. For this figure,
only the interaction terms were perturbed. The efficiency of the
dimeric core complex is shown for three orientations of the RCs (cf.
figure \ref{fig:DimerRotateRCs}): for the optimal orientation of
$\Phi_{RC} = 11^\circ$, for slightly misaligned RCs ($\Phi_{RC} =
21^\circ$), where the dimer has about the same $\Sigma$ as two
monomers, and for the most unfavorable configuration of $\Phi_{RC} =
110^\circ$.

The efficiency of the monomer is only slightly affected by the
disorder, it decreases from 0.15 without disorder to about 0.13 at
$\Delta E/E$ = 12\%. In the dimer, the advantage with optimally
aligned RCs over a core complex with slightly misaligned RCs vanishes
at already small perturbations, but at all perturbations the dimer is
more efficient than the monomer, as long as the RCs point into about
the right direction. For even stronger disorder, the efficiency of all
configurations, i.e, of the monomer and of the dimer with any
orientation of the RCs, tends to the same value of around 0.13.

Obviously, the monomer with its closed LH1 ring is more stable against
disorder than the open ring dimer. In other words, the closed LH1 can
easily be deformed away from its circular shape or the RC can move
inside the ring without degrading its antenna function noticeably. The
dimer, however, which is a more sophisticated and optimized structure,
has to be kept in shape in order to take advantage of its better
performance.

Experimental observations show, as we will explain later, that the
bacteria go the most obvious way to stabilize their dimeric LH1
antennae via a strong association bweteenm the flexible LH1 chain and
the globular RC.

\subsection{Open monomeric core complexes}

In the model, the spectrum of the empty closed monomeric LH1 has one
absorption line at 875 nm from two degenerate states (see above). When
a part of the LH1 ring is removed, the circular symmetry is broken.
Together with the now also nonsymmetric groundstate, the open LH1 ring
has three main absorbing states, the energies of which increase with
decreasing number of Bchls (data not shown). At $N=24$ Bchls, the
highest of these three levels comes in resonance with the RC
groundstate at 865 nm.

\begin{figure}[t]
	\begin{center}
		\includegraphics[]{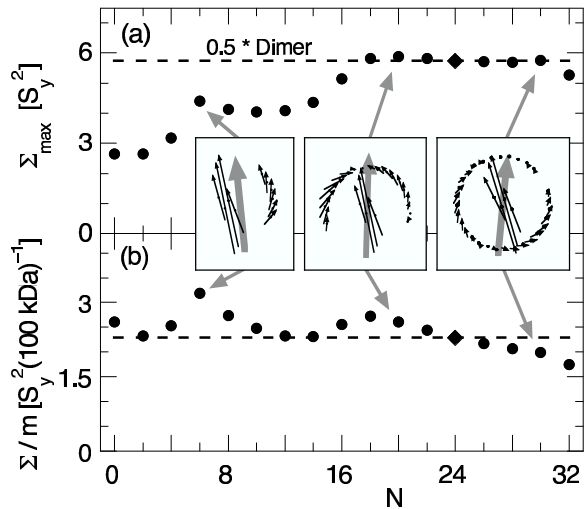}
	\end{center}
	\caption{Photosynthetic cross section $\Sigma$ for monomeric core
	complexes with partial LH1 rings with various numbers of Bchls,
	$N$: (a) $\Sigma$ for optimal orientation of the RC in the
	respective partial ring. (b) $\Sigma$ normalized to the total mass
	of an RC plus the partial LH1 ring. The RC contributes a mass of
	101 kDa and the complete LH1 ring of 200 kDa \cite{KOE96}. Half of
	the respective values from the dimer are indicated in both panels
	by the diamonds and the horizontal broken lines. The insets sketch
	the main photosynthetically active states at $N$=6, 20, and 30
	analogous to figure \ref{fig:DimerDipoles}.}
	\label{fig:partialSigma}
\end{figure}

In the open monomer, the orientation of the RC again determines
$\Sigma$. Panel (a) of figure \ref{fig:partialSigma} plots the maximal
$\Sigma$ at optimal orientation of the RC for core complexes from
$N=32$, i.e., for the closed monomer, down to $N=0$, which is an RC
without any LH1 chain. For comparison, half of the cross section of
the dimer is also given. One can discern three regimes: any
configuration in the range $18 \leq N \leq 30$, which is between a bit
more than a half ring and a nearly complete ring, has essentially the
same $\Sigma$, though for every $N$ the RC has a different optimal
orientation with respect to the symmetry axis of the partial ring.
This constant cross section is even higher than with the closed LH1
ring. The other two regimes are the ranges $6 \leq N\leq 14$ and $N
\leq 4$, i.e., from a quarter to a half ring, and the RC sided by just
a small part of the LH1. The reason for this behavior is that in each
of these regimes only a part of the LH1 chain couples directly to the
special pair Bchls of the RC. When the RC is aligned correctly, these
Bchls that do not contribute to the photosynthetically active states,
can be removed without degrading the performance.

A different kind of efficiency is plotted in panel (b) of figure
\ref{fig:partialSigma}. Here $\Sigma$ is not normalized to the number
of Bchls, but to the total mass of the core complex. According to
\cite{KOE96}, the RC has a mass of 101 kDa, while a complete LH1 ring
weighs 200 kDa. With respect to the total mass, the core complex with
the closed LH1 is the most inefficient configuration, while with any
partial LH1 the bacterium has to produce less material for the same
yield from direct photon absorption into the RC --- if the RC is 
oriented correctly.

Here again, to make use of the more sophisticated antenna, the
orientation of the RC has to be fixed relative to the LH1 and the LH1
has to be stabilized.

\subsection{Arrays of core complexes on a vesicle}

In a real bacterium there are many core complexes, sitting close
together. We therefore have to look at the efficiency of multiple
coupled core complexes.

\begin{figure}[t]
	\begin{center}
		\includegraphics[]{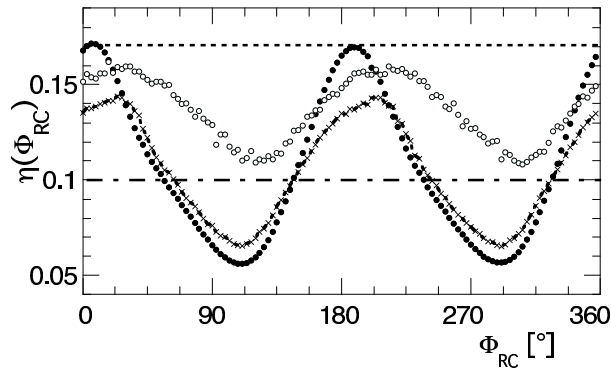}
	\end{center}
	\caption{Photosynthetic efficiency $\eta$ of an array of eleven
	dimeric core complexes on a vesicle vs. $\Phi_{RC}$ for different
	thermal fluctuations of the interactions of $\Delta E/E = 0$
	(solid dots), 1\% (crosses), and 8\% (open dots). The broken lines
	at $\eta = 0.17$ and at $\eta = 0.1$ indicate the efficiencies of
	a single thermally perturbed dimeric core complex (cf. figure
	\ref{fig:Rauschen}) and of an array of 24 monomeric core complexes
	with random orientations of their RCs, respectively. The
	three-dimensional geometries of the arrays are shown in figure
	\ref{fig:OnTheVesicle}.}
	\label{fig:etaArray}
\end{figure}

In the array of dimeric core complexes on a vesicle (see figure
\ref{fig:OnTheVesicle}), without any perturbation the most efficient
$\Phi_{RC}$ is the same as for one independent dimer and $\Sigma$
follows a similar curve, see figure \ref{fig:etaArray}. However, on
the vesicle the maximal $\eta$ is reduced from 0.21 to 0.17.
When this highly symmetric model array is perturbed, the behavior is
different from that of the isolated core complex. There, the
efficiency degraded monotonically with increasing fluctuations. In the
array, however, $\eta$ strongly decreases even with very small
perturbations of the interactions of $\Delta E/E \leq 1$\%, only to
increase again with increasing $\Delta E/E$. For strong perturbations,
the efficiency of the array on the vesicle comes close to that of an
equally perturbed isolated dimer, which is indicated in figure
\ref{fig:etaArray} by the broken line at $\eta=0.17$.

\begin{figure}[t]
	\begin{center}
		\includegraphics[scale=0.9]{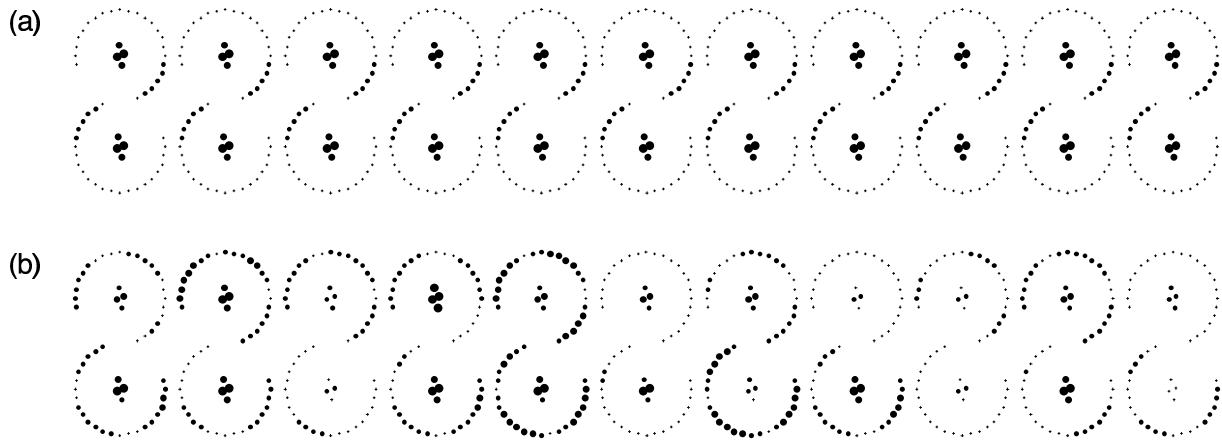}
	\end{center}
	\caption{Two eigenstates of an array of eleven dimeric core
	complexes on a vesicle, unrolled into the paper plane (cf. figure
	\ref{fig:OnTheVesicle}). The contributions of the individual
	dipoles, $|a_{ni}|^2$, are indicated by the size of the dots.
	$\Phi_{RC} = 25^\circ$ for both states. The upper panel (a) shows
	the main photosynthetically effective state of the unperturbed
	array with its perfect symmetry. Panel (b) shows a typical
	photosynthetically effective state at a thermal perturbation of
	the interactions of $\Delta E/E = 1$\% (cf. figure
	\ref{fig:etaArray}), which is localized over a few core
	complexes.}
	\label{fig:arrayStates}
\end{figure}

This behavior can be understood by comparing the eigenstates of the
unperturbed to these of the perturbed system. Two representative
states are sketched in figure \ref{fig:arrayStates}, one without and
one with fluctuations. Without fluctuations (panel (a)), the states
are highly symmetric. With fluctuations (panel (b)), the long range
order breaks down and the states become localized over three to five
core complexes. Seemingly, the symmetry of the unperturbed array,
which is reflected in the eigenstates, only allows for states with a
smaller efficiency than possible in an isolated dimer. The symmetry
breaking due to the thermal fluctuations relaxes this constraint, and
with a stronger perturbation the more efficient localized states lead
to the observed increase of the overall efficiency. \emph{In vivo},
the array of core complex dimers on a vesicle will never be perfectly
symmetric, as the vesicles are not rigid spheres. Consequently,
\emph{in vivo} no strict efficiency degrading long range order will
develop and the fluctuations limit the coupling between the core
complexes to their respective neighbours, which is good from the
perspective of overall efficiency.

Stochastic simulations showed that when a few core complexes are
coupled, at intermediate light intensities their yield is increased by
some 20\%, because the larger combined antenna reduces the statistical
fluctuations in the photon supply for each of the involved RCs
\cite{GEY06c}. This might explain, why it is advantageous for the
bacteria to closely pack the dimeric core complexes onto
chromatophores, even when this slightly reduces the efficiency of
absorption directly into the RCs.

As the monomeric core complexes are round and thus have no preferred
orientation of the RCs, a random distribution of $\Phi_{RC}$ was used
when placing the monomers onto the vesicle. Then, the efficiency of
the array of $\eta = 0.1$ is much smaller than for isolated core
complexes and essentially unaffected by the perturbations. This means
that the random orientation of the RCs already introduced more
disorder than the thermal fluctuations used here. Also one can
conclude that monomeric core complexes are better not packed onto
small vesicles, but spread out onto flat membranes, so that they do
not disturb each other.

Interestingly, up to now chromatophore vesicles were only found in
PufX expressing bacteria with dimeric core complexes, but not in
those, where the wild type has monomeric core complexes.

\section{Summary and conclusions}

Comparing the calculated photosynthetic cross sections and
efficiencies of the various configurations of core complexes --- the
monomer, the dimer, the open monomer, and the two arrays on a vesicle
built from monomers and dimers --- the following overall picture
emerges: without (thermal) perturbation and when the RC is oriented
optimally, the photosynthetic efficiency of the dimeric core complex
is about 30\% higher than that of the monomer. The orientation of the
RCs for maximal efficiency determined from the calculations, nicely
corresponds to their experimentally determined orientation
\cite{QIA05}. The open monomer is also more efficient than the closed
monomer and here, too, the orientation of the RC for maximal
efficiency reproduces the experimentally found orientation in
\emph{Rps. palustris} \cite{ROS03}. When the photosynthetic cross
section is related to the total protein mass of the RC plus the
partial LH1 ring, then any open configuration is more efficient than
the closed monomer. However, the closed monomeric core complex is the
most prominent form in purple bacteria.

With thermal fluctuations, the dimer remains more efficient than the
monomer, but with a smaller advantage over the monomer. Also it is
more sensitive to these perturbations. The thermal fluctuations were
modelled as quasistatic, i.e., as much slower than the photon
absorption event itself. Consequently, the closed monomer is also much
less sensitive to static deformations of the LH1 ring or to a
displacement of the RC away from its optimal position. Thus, the
monomer is the more flexible but less efficient configuration, while
the more efficient dimer has stronger requirements with regard to
structural stability.

The same trend --- that the dimer is more efficient, but easier
perturbed --- is found again, when the core complexes are put onto a
typical vesicle. Interestingly, for both the monomer and the dimer,
their efficiencies are smaller, when multiple identical core complexes
are coupled symmetrically. In this scenario, photosynthesis benefits
from the fluctuations, as they destroy the long range order on the
vesicle and lead to more efficient localized eigenstates.

Actually, there is a handful of experimental observations that fit
nicely with these findings: (i) AFM images of monomeric LH1 rings
without an RC showed that the LH1 rings themselves are quite flexible
and can easily be deformed \cite{BAH04b}. (ii) When dimeric core
complexes from \emph{Rb. sphaeroides} were reconstituted into planar
membranes, a quasicrystalline corrugated long range pattern developed,
which is best explained with rigid bent core complexes \cite{SCH04a,
GEY06a}. (iii) Dynamic experiments by, e.g., Barz \etal{} showed that
a mutation of \emph{Rb. sphaeroides}, where only the expression of
PufX is suppressed without further modification of the LH1, so
strongly slows down the diffusion of the quinones to and from the RC,
that this PufX$^-$ mutant can not live on photosynthesis any more.
However, its photosynthetic competence is partly restored, when the
$\alpha$ or $\beta$ subunits of the LH1 are modified, too
\cite{BAR95a, BAR95b}. (iv) For \emph{Rps. rubrum}, a species without
PufX, a recent calculation estimated that a quinone molecule can pass
the LH1 ring within about one millisecond, which is fast enough to not
impede photosynthesis \cite{AIR06}. (v) In the latest high resolution
EM images of the dimeric core complex from \emph{Rb. sphaeroides}, two
tentative positions of PufX were identified. It either sits at the
joint between the two LH1 halves or between the open ends of the LH1
chains and the RCs \cite{QIA05}.

From our calculations and these observations, we put forth the
hypothesis, that in these bacteria that express PufX an additional
modification of the LH1 chain leads to a strong association between
the LH1 and the RC. This would explain, why the dimeric core complexes
are rigid, even if the LH1 itself is floppy. In PufX$^+$ species the
LH1 chain would stick to the nearly globular RC and thus be
stabilized, while in species without PufX in the wildtype the RC can
float inside the easily deformed, but closed LH1. Loosely placed
inside the closed monomeric LH1 ring, the RC will rotate, but this,
too, has no effect on its function. Actually, when the LH1 is easily
deformed and there were no association between RC and LH1, the RC
would diffuse out of an open LH1. Thus, for the core complex to be
stable with an open LH1, these two have to stick together rather
strongly.

With regard to the orientation of the RC, it is interesting that Qian
\etal{} \cite{QIA05} identified a putative position of the PufX
between the open ends of the LH1 ring and the long side of the RC. At
this position, PufX could fix the open end of the LH1 chain to the RC
and also lock the orientation of the RC with respect to the gap in the
LH1 chain. A similar explanation would apply to the position of the
PufX homolog found in \emph{Rps. palustris}, which sits between one
end of the open monomeric LH1 and the long side of the RC
\cite{ROS03}.

From the experiments of Barz \etal{} \cite{BAR95a, BAR95b} it became
clear that PufX is required for fast quinone exchange at the RC. If in
\emph{Rb. sphaeroides} the LH1 sticks tightly to the RC and in its
PufX$^-$ mutant the gap in the LH1 is missing, then the quinone
binding site is blocked. In the suppressor mutants with the modified
LH1 chain, the association between the RCs and the LH1 would be
weaker, which would allow the quinones to reach the RCs much faster,
partly restoring photosynthetic growth.

Consequently, the gap in the LH1 dimer has a dual function: it is both
a part of the design of a more efficient antenna, stabilized by PufX
and a strong association between the LH1 and the RCs, and it allows
for an even faster exchange of the quinones than in species with a
loosely attached, but closed LH1 ring.


\end{document}